\documentclass[useAMS,usenatbib]{mn2e}
\usepackage{epsfig}
\usepackage{float}
\usepackage{times}
\usepackage{longtable}
\usepackage{float}
\usepackage{graphicx} 
\usepackage{epstopdf}
\usepackage{amsmath}
\usepackage{threeparttable }
\begin{document}

\title[Radio transients] 
{The variability timescales and brightness temperatures of radio flares from stars to supermassive black holes}
 
\author[M. Pietka, R.P. Fender and E.F. Keane]
       {M. Pietka$^{1,2}$\thanks{email: malgorzata.pietka@astro.ox.ac.uk}, R.P. Fender$^{1,2}$, E.F. Keane$^{3,4}$\\ 
$^1$ Astrophysics, Department of Physics, University of Oxford, Keble Road, Oxford OX1 3RH, UK\\
$^2$ Physics and Astronomy, University of Southampton, Highfield, Southampton, SO17 1BJ, UK\\
$^3$ Centre for Astrophysics and Supercomputing, Swinburne University of Technology, Mail H30, PO Box 218, Hawthorn, VIC 3122, Australia\\
$^4$ ARC Centre of Excellence for All-sky Astrophysics (CAASTRO).
}

\maketitle
\begin{abstract}
In this paper we compile the analysis of $\sim 200$ synchrotron flare events from $\sim 90$ distinct objects/events for which the distance is well established, and hence the peak luminosity can be accurately estimated. For each event we measure this peak and compare it to the rise and decay timescales, as fit by exponential functions, which allows us in turn to estimate a minimum brightness temperature for all the events. The astrophysical objects from which the flares originate vary from flare stars to supermassive black holes in active galactic nuclei, and include both repeating phenomena and single cataclysmic events (such as supernovae and gamma ray burst afterglows). The measured timescales vary from minutes to longer than years, and the peak radio luminosities range over 22 orders of magnitude. Despite very different underlying phenomena, including relativistic and non-relativistic regimes, and highly collimated versus isotropic phenomena, we find a broad correlation between peak radio luminosity and rise/decay timescales, approximately of the form $L \propto \tau^{5}$. This rather unexpectedly demonstrates that the estimated minimum brightness temperature, when based upon variability timescales, and with no attempt to correct for relativistic boosting, is a strongly rising function of source luminosity. It furthermore demonstrates that variability timescales could be used as an early diagnostic of source class in future radio transient surveys. As an illustration of radio transients parameter space, we compare the synchrotron events with coherent bursts at higher brightness temperatures to illustrate which regions of radio transient parameter space have been explored.
\end{abstract}

\section{Introduction}
\label{introduction}

It has been known for 50 years that synchrotron-emitting radio sources can vary dramatically, undergoing flaring events. These events are associated with particle acceleration and/or compression of magnetic fields during phases of rapid injection of kinetic energy into the ambient medium. Sources from the stellar scale to supermassive black holes, a range of over $10^9$ in mass, can exhibit this behaviour, often associated with large amplitude variations at other (e.g. optical, X-ray) wavelengths. Radio emission can therefore be used as a measure of the degree of kinetic feedback associated with a given phase or phenomenon, and also as an identifier and locator of interesting events for study at other wavelengths. The kinetic feedback processes associated with this feedback are not a homogeneous set, however, and cover a wider range of geometries and physical regimes. In accreting black holes (of all masses, stellar to supermassive) and neutron stars, as well as gamma-ray burst afterglows (which are likely to be one of the former) the radio emission is often associated with highly collimated (opening angles of degree scale or smaller) flows with significant bulk Lorentz factors (at least in the initial phases). In novae and supernovae, however, the feedback is in the form of much more slowly moving (1000s of km s$^{-1}$ or less) ejecta moving in a much more isotropic geometry. Intermediate geometries and regimes also exist. 

One of the earliest models for these radio flares was that of van der Laan (1966) in which a cloud of relativistic particles and magnetic field expands adiabatically into a surrounding medium. In this model two competing effects determine the light curve at a given frequency as the cloud expands: decreasing optical depth causes an increase in the flux, but adiabatic expansion losses reduce the internal energy and therefore the synchrotron emission. Because the optical depth to synchrotron self absorption decreases with increasing frequency, higher frequencies peak earlier and stronger, although in the later, optically thin, phase the lowest frequencies are strongest. Van der Laan (1966; and many others since) clearly demonstrates (his Fig. 1) the contrast between the sharply peaked light curves at high radio frequencies and the much smoother events at lower frequencies.

Many modifications have been made to this simple model, including internal and external shocks at later times which can re-energise particles, constrained expansion of the ejecta, deceleration from highly to mildly relativistic bulk flows and, as noted above, highly variable geometry depending on the originating astrophysical phenomena. Examples of such models include gamma-ray burst (GRB) afterglow ( Chevalier \& Li 1999; Li \& Chevalier 1999; Sari, Piran \& Halpern 1999; Kumar \& Panaitescu 2000), Supernovae (SN, Chevalier 1982a,b; Weiler et al. 2002), tidal disruption event (TDE, Giannios \& Metzger 2011), jets in X-ray binaries (XRB) and active galactic nuclei (AGN) (Hjellming 
\& Johnston 1988; Livio 1999; Fender 2006).

We are currently in an era in which wide-field blind surveys for radio transients are beginning (e.g. with LOFAR: van Haarlem et al. 2013; Fender \& Bell 2011 for a general overview) and are going to be a major factor for future larger radio facilities such as MeerKAT (Booth et al. 2009), ASKAP (Johnston et al. 2007, 2008) and the Square Kilometre Array (SKA; Fender et al. 2015 in prep). As in the X-ray and optical fields, in which such blind searches for transients have been undertaken for some time (e.g. Sesar et al. 2007), early identification and classification of a new variable or transient, even if crude, can be invaluable in optimising the science delivered from a finite set of resources (observing time). The simplest and earliest such diagnostic is the variability timescale of the event (see e.g. Rau et al. 2009, Ofek et al. 2014, Cao et al. 2012 for the timescale-magnitude plot for optical transients used in PTF literature).

There is no strong {\em a priori} reason to believe that a simple relation should exist between event type and the most basic characteristics of radio variability, given the wide range of geometries and physical regimes present, as outlined above. In this paper we compare the peak radio luminosities, rise and decay timescales of radio flare events across the widest possible range, and find that a more or less monotonic relation does exist, albeit with some scatter. This has implications for the intrinsic, basic, similarity of the different models (which are, it should be stressed, successful in fitting events in a single class). Furthermore, the steepness of the observed relation between luminosity and timescales implies, rather unexpectedly, that the minimum brightness temperature, of a given event is an increasing function of source luminosity. Finally, the result offers hope that the relation could be used as a very early time diagnostic of radio flaring events and transients.

\section{Data analysis}
\label{Data}

The following section presents the analysis performed on a sample of $\sim$ 200 light curves compiled from the literature and the Green Bank Intereferometer (GBI) archive \footnote{ftp://ftp.gb.nrao.edu/pub/fghigo/gbidata/gdata/}. 
Subsections below describe the data sample and details of the analysis.

\begin{figure}
\includegraphics[scale=0.301]{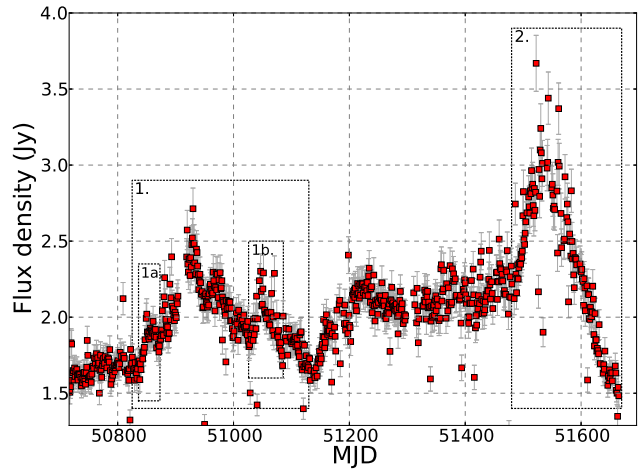}
\caption{An example of light-curve (blazar 0851+202; GBI, 8.3 GHz) from which several flares have been selected for the analysis. 
Apart from two main long time-scale flares ({\em1.,2.}), we have also included two shorter flares that took place during rise ({\em 1a.}) and decline ({\em 1b.}) phases of the first outburst. Selection of all the flare events has been done by eye.}
\label{Fig1}
\end{figure}

\begin{figure}
\includegraphics[scale=0.25]{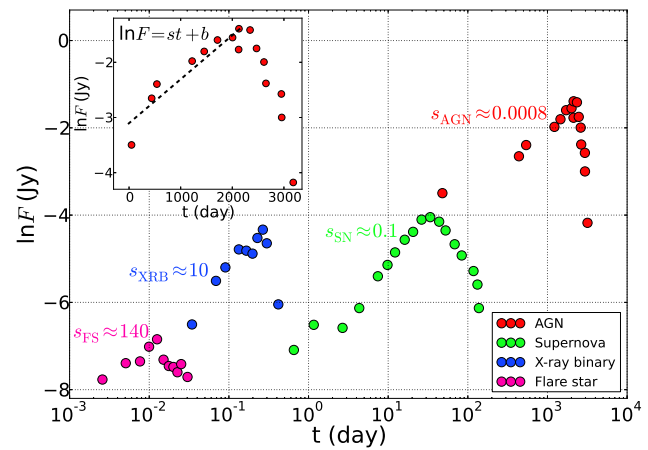}
\caption{Examples of the light-curves for several classes of objects, illustrating different time-scales of the events: flare star (AU Mic), X-ray binary (Sco X-1), supernova (SN 1994I) and AGN (NGC 4278).
Schematical example of the slope measurement of the light curve for rising phase is shown for the flare from NGC 4278. The measured slope $s$ of the fitted line corresponds to $ \tau^{-1}$, where ${\tau}$ is the e-folding time. 
All the light-curves shown have been normalized to have a similar initial flux density.}
\label{lightcurves}
\end{figure}

\subsection{Data}
\label{sec:sample}

\begin{figure*}
\includegraphics[width = \textwidth]{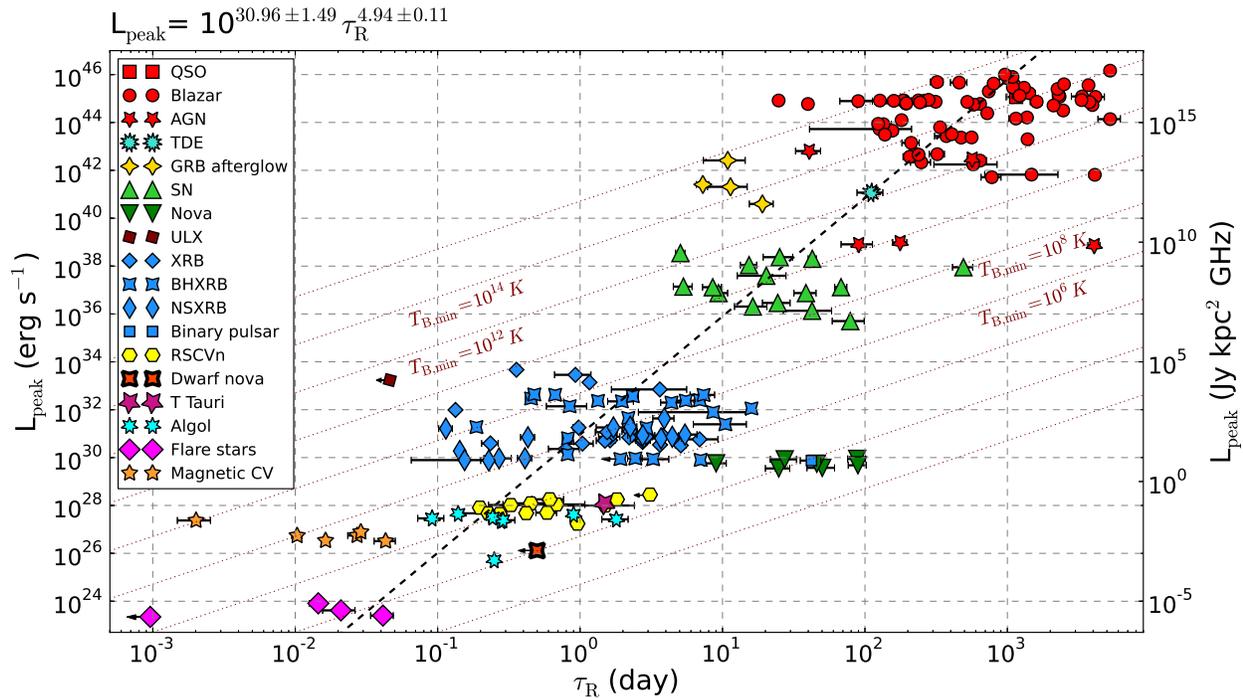}
\caption{Exponential rise timescale as a function of peak radio luminosity for the entire set of radio flares studied. The overall relation is of the form $L \propto \tau^{4.94\pm0.11}$.
Overplotted are lines corresponding to a fixed minimum brightness temperature ($T_{\mathrm{B, min}}$), under the assumption that the size of the emitting region $r = c \tau$, which have a form $L \propto \tau^2$, demonstrating clearly that $T_{\mathrm{B, min}}$ is an increasing function of luminosity, peaking just above the theoretical limit of $10^{12}$K for the most luminous AGN (which are also likely to be beamed).}
\label{rise}
\end{figure*}

In order to investigate the relation between variability time scales and peak radio luminosities, we compiled a data set encompassing as wide a range of parameters as possible, across several different classes of object.

The data consists of light curves corresponding to single outbursts, that were available in the literature and the GBI database. Crucially, the distances of sources in the presented sample and hence peak radio luminosities are known. Additionally, due to the strong frequency dependence of the shape of the light curve (Section \ref{introduction}), we limited our sample to sources that were observed at 5 - 8 GHz. In particular, events observed in the MHz band with e.g. LOFAR, MWA will have much smoother ("less peaked") curves which will both deviate from the relation we present here and be harder to detect automatically in software in future blind searches.

All the analysed sources together with the references are listed in Tables \ref{agnsample} -- \ref{fssample}. For objects that undergo recurrent outbursts (XRBs, AGN, etc.) we analysed multiple flare light curves, if such were available.
Selection of flares used in the analysis has been done by eye. We included events with the best possible time coverage, for which the identification of rise/decline phases was 	possible. For objects with multiple flares, measurements have been done on different time scales, including both longer timescale variability and shorter outbursts. Examples of selecting flares from the noisy data are shown in Figure \ref{Fig1}.

Note that we do not include any type of intermittent or flaring pulsar (Lyne 2010), or Fast Radio Burst (Lorimer et al. 2007; Thornton et al. 2013 and references therein) in our initial analysis, as they are {\em coherent} events with far higher brightness temperatures than possible for synchrotron emission (Readhead 1994). However, we do consider such events later in our analysis (Section \ref{PSF}).
Data points of the light curves selected from the literature have been extracted for the further analysis using WebPlotDigitizer \footnote{http://arohatgi.infobsai/WebPlotDigitizer/} software.

\subsection{Method and Analysis}
\label{dim}

To examine the relation between the radio luminosity of the source and the variability time scale, we split the light-curve into rising and declining phases, and consider them separately. Figure \ref{lightcurves} shows examples of light-curves for several classes of object and the measurement method. A linear fit was made to the natural logarithm of the flux as a function of (linear) time; in other words we fit exponential functions.
These have the advantage of being distance independent and (in some models) physically motivated. In the van der Laan model (1966), the radio flux density of a source is described by an exponential function that depends on the optical depth. As the optical depth will decrease as the size of the source increases over time, the flux density will be seen to change  exponentially. It is true that some alternative models favour power-law decays. These power-law fits (linear fits to log flux vs log time) were also attempted but exponential fits gave the most robust and reproducible results.

Slope measurements were made on a case by case basis in {\em Mathematica} once the start and end points of an event had been chosen by eye. The steps described above were applied to the sample of light curves presented in Section \ref{sec:sample}.
Peak radio luminosities have been calculated using 

\begin{equation}
L_{\mathrm{peak}}= \nu  L_{\nu,\mathrm{peak}} = 4 \pi d^{2}  \nu F_{\mathrm{max}} ,
\end{equation}

\noindent where  $L_{\nu,\mathrm{peak}}$ is the peak monochromatic luminosity, $\nu$  is the observed frequency, $F_{\mathrm{max}}$ is the  maximum flux density and d is the luminosity distance to the source. In other words we make no attempt to correct for any beaming or anisotropy.

\section{Variability time scales for different classes of radio sources}
\label{timescales}

Figure \ref{rise} shows the results of the measurements described in Section \ref{Data} for a sample of 200 (200 for rise, 200 for decline phases) light curves of 70 sources found in the literature (mostly observed at $\approx$ 5 GHz) plus 27 objects from the GBI database (8.3 GHz) (out of which light curves for 5 objects analysed both from the literature and GBI). The variability time scale parameters $\tau$ calculated for rising phases of light-curves in the sample are plotted against the corresponding peak radio luminosity of the source. We see that as the peak luminosity increases, sources tend to vary on longer time scales. This is not unexpected (see Introduction and Discussion) but has not been systematically measured before. AGN as the most luminous objects in the Universe can take up to several years before they reach peak flux density, while nearby and faint Flare Stars go into outburst and fade away within a couple of hours. A very similar trend is followed by the decline rate (not plotted here). The analysed sample of objects includes synchrotron emitting sources as well as those for which the origin of radio emission is gyro-synchrotron (flare star, magnetic CV, Algol and RSCVn). This different flavour of incoherent emission is explained by radiation from electrons trapped in magnetic coronal loops. Table \ref{fits} shows the results of fitted values for the correlation between $L$ and $\tau$ for two cases: first, including all of the sources in our sample and second - excluding gyro-synchrotron emitting sources.

\begin{table}
\begin{threeparttable}[b]
\begin{tabular}{|l|@{}c|@{}c|c|@{}c}
\hline
& a $\pm \delta$ a & b $\pm \delta$ b & a\ensuremath{'} $\pm \delta$ a\ensuremath{'}   \tnote{1} & b\ensuremath{'} $\pm \delta$ b\ensuremath{'}  \tnote{1}\\
\hline 
Rise        & 4.94 $\pm$ 0.11 & 30.96 $\pm$ 1.49 & 4.84 $\pm$ 0.16 & 31.19 $\pm$ 2.39\\
Decline & 4.91 $\pm$ 0.20 & 29.46 $\pm$ 2.66 & 5.48 $\pm$ 0.25 & 28.09 $\pm$ 2.99\\
\hline
\end{tabular}
\begin{tablenotes}\footnotesize
\item [1] Additional fit done for comparison on a sample excluding flares that originate from gyro-synchrotron emission mechanism (flare star, magnetic CV, algol, RSCVn). 
\end{tablenotes}
\end{threeparttable}
\caption{Parameters of the fits to the rise/decline phases. Fits done with  {\em Mathematica}, with the following formula: $\log(L_{\mathrm{peak}}) = a \times \log\tau + b$.}
\label{fits}
\end{table}

\begin{figure}
\includegraphics[scale=0.56]{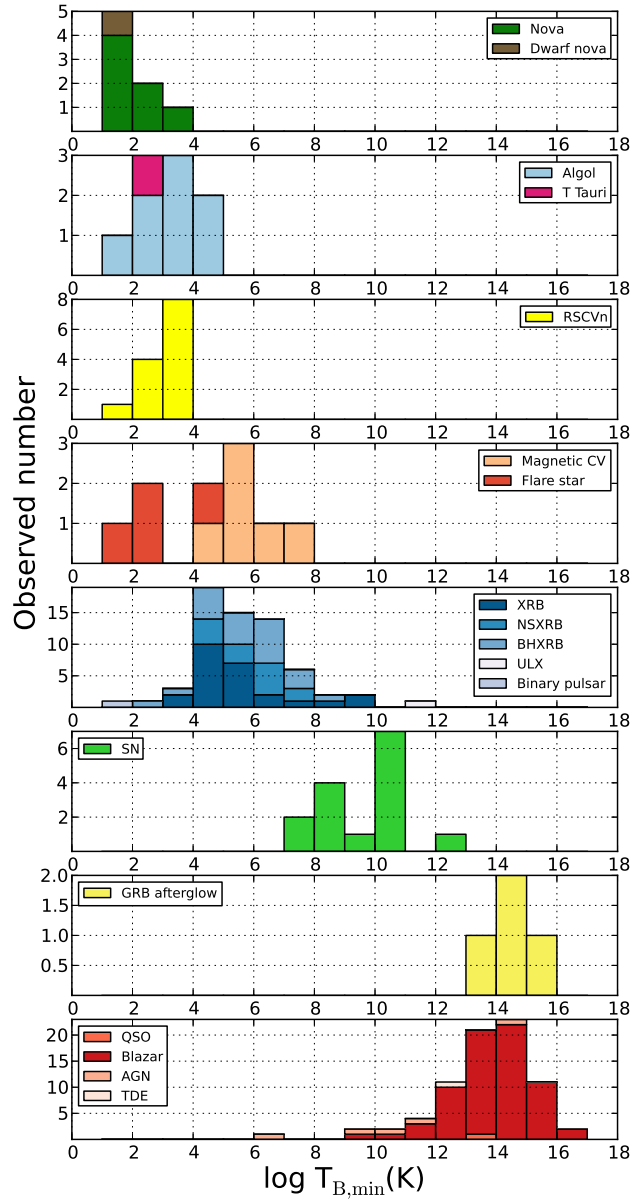}
\caption{The distribution of minimum brightness temperature, $T_{\mathrm{B,min}}$ for the flare events studied.}
\label{bt} 	
\end{figure}

\section{Brightness temperature}
\label{BT}

In a steady state, the brightness temperature of synchrotron emission is limited to $T_{\mathrm{B}} \sim 10^{12}K$ (Readhead 1994); above this value inverse Compton losses rapidly cool the electrons. The combination of luminosity and timescale naturally allow us to constrain the effective $T_{\mathrm{B}}$ for each flare event measured in this study.

In the Rayleigh-Jeans part of the blackbody spectrum, the monochromatic luminosity of a spherical source of radius $r$

\begin{equation}
L_{\nu} =\frac{2 k_B T \nu^2}{c^2} 4 \pi r^2 .
\end{equation}

\noindent A minimum brightness temperature, $T_{\mathrm{B,min}}$ will correspond to the source having a maximum size $c \tau$ where $\tau$ is the variability timescale. Substituting and rearranging, we find that

\begin{equation}
T_{\mathrm{B,min}} =\frac{\Delta L_{\nu}}{8 \pi k_B \nu^2 \tau^2},
\end{equation}

\noindent which allows us to calculate $T_{\mathrm{B,min}}$ for any source from which we have measured the luminosity and variability timescale (note that this does not take into account relativistic beaming). We note that the luminosity $L_{\nu}$ used above should be a fraction of the monochromatic peak luminosity $L_{\nu, \mathrm{peak}}$, since $\tau$ is the exponential rise (or decay) timescale:

\begin{equation}
\Delta L_{\nu} = \frac{e-1}{e} L_{\nu,\mathrm{peak}}   .
\end{equation}

\noindent Using this approach we calculate the $T_{\mathrm{B,min}}$ for all the events in our sample, and present a histogram of the results in Fig \ref{bt}.

It is interesting to note that if all the sources had the same brightness temperature and relation between variability timescale and physical size, we would expect a relation between luminosity and timescale of the form 

\begin{equation}
L \propto \tau^2,
\end{equation}

\noindent which is considerably shallower than the relation we observe (Fig \ref{rise}), which is approximately

\begin{equation}
L \propto \tau^{5} .
\label{rule}
\end{equation}

\noindent This is illustrated by the lines of constant $T_{\mathrm{B,min}}$ overplotted in Fig \ref{rise}. 
This is an unexpected result which we discuss in  Section \ref{discussion}.

\begin{figure*}
\includegraphics[width = \textwidth]{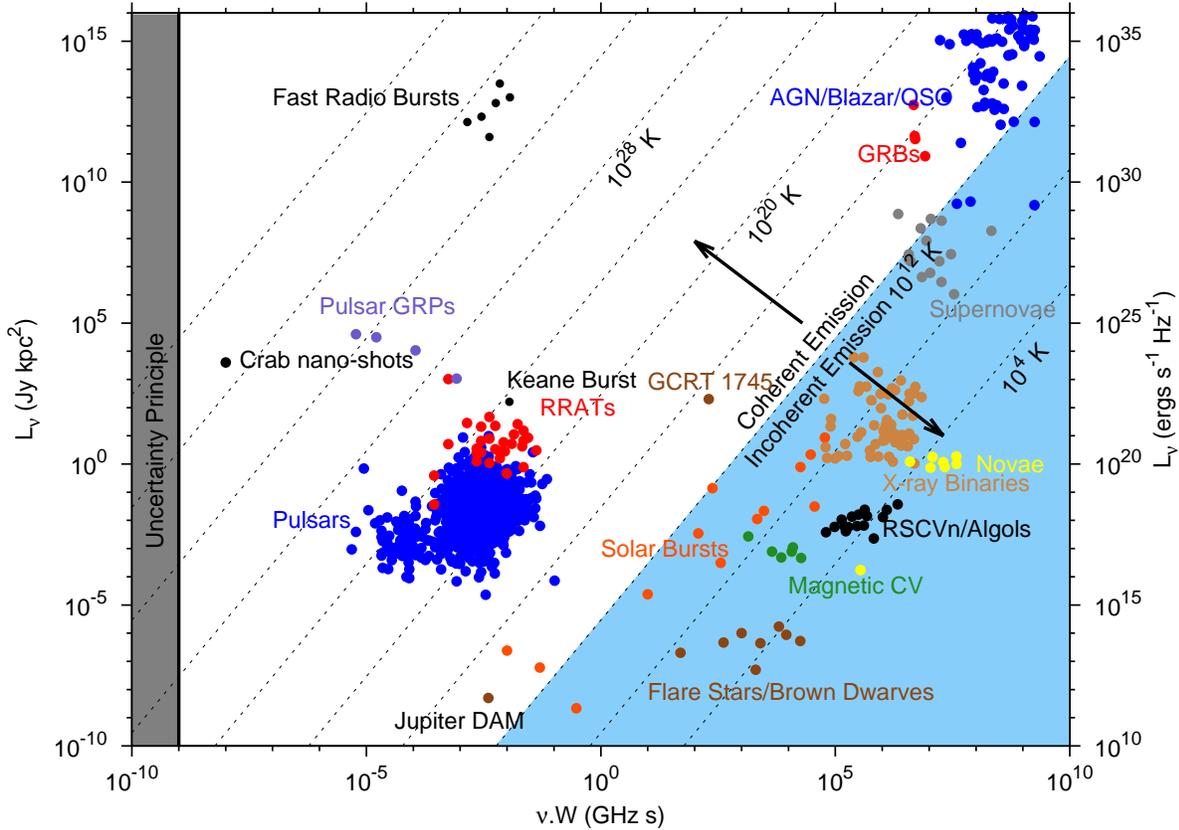}
\caption{Plotted is a variation of the Figure 2, with $L_{\nu}$ plotted against
$\nu W$ over an even wider range of parameter space. This allows us to
identify the sources of coherent radio emission (pulsars, fast radio
bursts, certain emission from Jupiter and the Sun, etc.) for
comparison with the more slowly varying synchrotron transients.  As can clearly be seen the correlation (Fig \ref{rise}, Table \ref{fits}, Equation \ref{rule}) applies only to the imaging/synchrotron sources, and cannot be used for classifying beamformed/coherent sources.}
\label{ps}
\end{figure*}

\section{Exploring Radio Transient Parameter Space}
\label{PSF}

Having identified the correlation (Equation \ref{rule}) between peak radio luminosity 
and the variability timescale of synchrotron transients, which applies over so many
orders of magnitude, it makes sense to examine how far this
relation can be stretched. To this end we examine the parameter space
of the fastest transients, from the millisecond bursts seen from most
pulsars and fast radio bursts, to the nanosecond shot emission seen
from the Crab pulsar (Keane 2013, Thornton et al. 2013). Transients are
often classified as either `beamformed', i.e. those detected in high
time-resolution pulsar-like observations where dedispersion is
required, or `imaging', i.e. sources can be identified in images and
dedispersion is not required (Hassall, Keane, \& Fender 2013). The boundary
between these classes is often taken to be $\sim 1$~s, so that the
synchrotron sources considered in Sections \ref{Data}, \ref{timescales}, \ref{BT} fall into the latter
category. The extended parameter space is shown in Figure \ref{ps}. We can
see that the relation we have determined does not
apply to beamformed transients, which are all examples of coherent
emission, but rather seems to be a manifestation of the physics
underlying incoherent synchrotron emission only.

\section{Discussion and Conclusions}
\label{discussion}

We have for the first time compiled the analysis of the simplest measurable quantity of a synchrotron flare event, its rise time, with its luminosity. It comes as no surprise that the most luminous events vary more slowly, as can be illustrated by the following example. The main energy loss mechanism for these phenomena is the work done in expanding against the surrounding medium, and yet the maximum 3D expansion speed of a relativistic plasma is fixed at $c / \sqrt{3}$. Since the maximum (rest frame) brightness temperature (surface brightness) of a synchrotron emitting plasma (in equilibrium) is $\sim 10^{12}$K, then the most luminous events are going to be associated with larger sources. The scale factor ($\rho = r / r_0$ in the original model of van der Laan 1966), on which the observed flux is strongly dependent (the flux rises as $\rho^3$ in van der Laan during the optically thick phase), is therefore going to evolve more slowly for the most luminous sources. In other words, a doubling in size will result -- in this simple model -- in the same relative flux change, but will take longer for more luminous and larger sources. Of course, as discussed in the introduction, since van der Laan there have been a large number of further, more complex models for different scenarios. It was therefore not at all obvious that a compilation such as this would result in the rather straightforward relation between rise/decay times and radio luminosity. 

\begin{figure}
\includegraphics[scale =0.325]{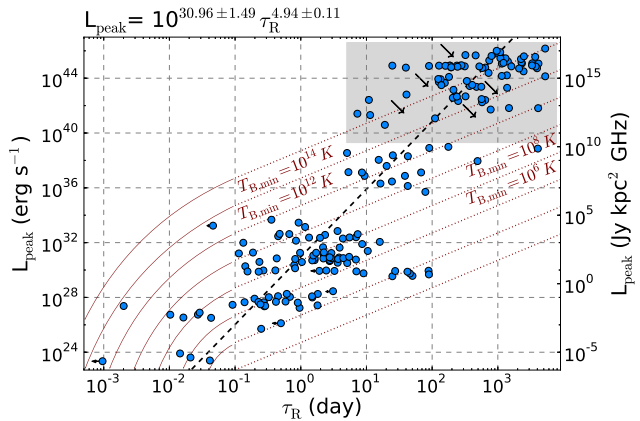}
\caption{This plot shows Figure \ref{rise} schematically modified to illustrate possible effects that could explain the observed increase in the estimated minimum brightness temperature with luminosity. Error bars and differentiation between classes have been removed for clarity. In the lower luminosity end of the sample, minimum brightness temperatures might be overestimated due to the smaller radius of the emitting region. 
Constant $T_{\mathrm{B,min}}$ lines plotted in Fig \ref{rise} would in fact need to turn down while approaching lower luminosity sources. On the other end of the plot the most luminous sources in the sample are expected to be relativistically beamed. The actual luminosity of these objects could be significantly lower, and the variability timescale slower than what we observe (marked with arrows). The likely result of both effects is to reduce the actual spread in the intrinsic values of $T_{\mathrm{B}}$.}
\label{Fig6}
\end{figure}

However, approaching the observed luminosities and times from the point of view of the brightness temperature, we find that the relation we find, $L \propto \tau^{5}$, is much steeper than that which would be expected for a homogeneous set of phenomena with the same brightness temperature, $L \propto \tau^2$. What is the origin of this effect? If intrinsic to the sources (and not due to relativistic beaming, see below), it could be perhaps due to a systematically varying relation between the true radius and the variability timescale, or to an increasing energy density with increasing luminosity. The former effect could mean that for a given timescale, the true radius of the emitting region is considerably smaller than $c \tau$, meaning we are severely underestimating the $T_{\mathrm{B}}$. This is the case for flare stars for example, where the Alfv\'{e}n speed should be considered rather than $c$. This would mean a velocity of $\times$10$^{6}$ ms$^{-1}$(Mitra-Kraev et al. 2005), which would change the estimated minimum brightness temperatures by factor of 10$^{4}$. 
Figure \ref{Fig6} schematically shows the shape of constant $T_{\mathrm{B,min}}$ lines we would expect if this effect was accounted for.

However, it is hard to see this explaining everything: since the sound speed of an ultrarelativistic plasma is $\sim c/\sqrt{3}$ and we believe that this condition should hold for both AGN and XRBs, both of which seem to have highly relativistic jet plasmas, it is not clear why they should be systematically two orders of magnitude different in $T_{\mathrm{B,min}}$ from each other. The actual geometry of relativistic jets is likely to be rather linear and so a 3D expansion at $c/\sqrt{3}$ is not likely, but could/should be similar for XRBs and AGN.
On the other hand, an increasing energy density with luminosity does not seem obvious, either, since the jets of AGN are anticipated to be {\em lower} density than those of XRBs (under simple assumptions, the energy density at a given Eddington ratio should vary as $M^{-1}$).

Can relativistic beaming explain the trend perhaps? This would result in a larger apparent $T_{\mathrm{B,min}}$, and it is certainly the case that blazars and GRB afterglows, which make up a large fraction of the luminous end of the sample, are pointed towards us. Such sources will appear to be brighter and vary faster than in their comoving frames, resulting in a higher apparent minimum brightness temperature.
This effect is schematically shown in Figure \ref{Fig6}, where arrows indicate the expected direction that the AGN sample would move if we could account accurately for Doppler boosting.
Therefore, in a flux-limited sample of such jetted sources you would expect to see an increase in $T_{\mathrm{B,min}}$ with luminosity.   
This is analagous to Malmquist bias, in which we are systematically seeing things which are increasingly beamed towards us as we observe them at larger distances. This effect has been investigated for AGN in some detail (e.g. Homan et al. 2006, Vermeulen \& Cohen 1994). However, it seems unlikely that this alone can explain the near-monotonic trend towards higher $T_{\mathrm{B,min}}$ with luminosity across our entire sample for at least two reasons: (i) some of the objects in the sample -- e.g. supernovae, flare stars -- are not expected to be significantly boosted, (ii) the lowest-luminosity population of relativistic sources -- the X-ray binaries -- are actually an {\em X-ray selected} sample which in general are de-boosted due to large angles to the line of sight.
The real origin of the trend is likely to be a combination of both effects, but we cannot rule out a dominant effect which we have not considered here.

Of course as well as being astrophysically interesting, the relation opens up the possibility of using variability timescale as an early classifier of radio transient events which may be discovered by future wide-field monitoring in the radio band (see Fender \& Bell 2011 and references therein). A faint (sub-mJy) event could be anything from a Flare Star to a Supernovae, and yet within minutes/hours (probably) and days (certainly) it would be obvious from the rise time which class of object it was likely to have been. However, the likelihood of a given event cannot be simply read from our figures, since these are compiled from highly biased samples. In a future work we plan to convolve the compiled data with estimates of the areal density (rates) of the different classes of objects, and incorporate these into an automated light curve classifier.

\section*{Acknowledgements}
The authors would like to thank James Miller-Jones, Magaretha Pretorius and Rachel Osten for helpful discussions.
This project was funded by European Research Council Advanced Grant 267697 "4 PI SKY: Extreme Astrophysics with Revolutionary Radio Telescopes". 
EFK acknowledges the support of the Australian Research Council Centre of Excellence for All-sky Astrophysics (CAASTRO), through project number CE110001020.
The Green Bank Interferometer is a facility of the National  Science Foundation operated by the NRAO in support of NASA High Energy Astrophysics programs. This research has made use of data from the MOJAVE database that is maintained by the MOJAVE team (Lister et al., 2009, AJ, 137, 3718). This research has made use of the NASA/IPAC Extragalactic Database (NED) which is operated by the Jet Propulsion Laboratory, California Institute of Technology, under contract with the National Aeronautics and Space Administration.

\section*{Appendix -- source tables and data}

\begin{table*}
\centering
\caption{Sample of AGN, blazars (including flat spectrum radio quasars) and a compact steep spectrum radio quasar (QSO) used in the analysis. }
\begin{threeparttable}[b]
\begin{tabular}{cccc}
\hline
Source & Distance (Mpc) &  Reference & Distance reference \\
\hline 
0224+671 & 2981 & GBI& MOJAVE\tnote{1 } / NED\tnote{2} \\
0235+164 & 6142 & \citet{0235+164} & MOJAVE / NED\\
0336-019 & 5434 & GBI& MOJAVE / NED\\
0458-020 & 18551 & \citet{Pyatunina2006} & MOJAVE / NED\\
0528+134 & 16418 & \citet{Pyatunina2006}, GBI & MOJAVE / NED\\
0605-085 & 5577 & \citet{0605-085} &MOJAVE / NED\\
0850-121 & 3281 & GBI& MOJAVE / NED\\
0851+202 & 1577 & GBI& MOJAVE / NED\\
0954+658 & 1951 & \citet{0954+658}, GBI & MOJAVE / NED\\
1237+049 & 13441 & GBI&MOJAVE / NED\\
1328+254 & 7093 & GBI& MOJAVE / NED\\
1413+135 & 1232 & GBI& MOJAVE / NED\\
1622-297 & 5142 &GBI& MOJAVE / NED\\
1749+096 & 1673 & GBI& MOJAVE / NED\\
1803+784 & 4107 & \citet{1803+784} & MOJAVE / NED\\
2004-447 & 1192 & \citet{2004-447} & \citet{2004-447}\\
2005+403 & 13205 & GBI& MOJAVE / NED\\
2223-052 & 10138 & \citet{2223-052} & MOJAVE / NED\\
\hline
3C120  & 143 &  &\\
3C273 & 748 & &\\
3C279 & 3071 & \citet{3C} & MOJAVE / NED\\
3C345 & 3473 && \\
3C454.3 & 5489 && \\
\hline
BL Lacertae & 305 & \citet{BLLac}, GBI & MOJAVE / NED\\
CTA 102 & 6943 & \citet{2223-052} & MOJAVE / NED\\
NGC 4278 & 14.9 &  \citet{NGC4278} &  \citet{NGC4278D} \\
NGC 7213& 22 & \citet{NGC7213} & \citet{NGC7213D} \\
NRAO 150 & 11193 &\citet{NRAO150} & MOJAVE / NED\\
NRAO 530 & 5834 & \citet{NRAO530}, \citet{Pyatunina2006} & MOJAVE / NED\\
Swift J1644+57 & 1845 & \citet{J1644+57} & \citet{swiftD} \\
III zw2 & 403 & \citet{IIIzw2} & MOJAVE / NED\\
\hline
\end{tabular}
\begin{tablenotes}\footnotesize
\item [1] http://www.physics.purdue.edu/astro/MOJAVE/index.html
\item [2] http://ned.ipac.caltech.edu/ 
\end{tablenotes}
\end{threeparttable}
\label{agnsample}
\end{table*}

\begin{table*}
\centering
\caption{Supernovae sample used in the analysis.}
\begin{tabular}{cccc}
\hline
Source & Distance (Mpc) &  Reference & Distance reference\\
\hline
SN 1980K & 5.9 & \citet{SN1980K} & \citet{SN1980KD}\\
SN 1988Z & 93 & \citet{SN1988Z} & \citet{SN1988ZD}\\
SN 1993J & 3.6 & \citet{SN1993J} & \citet{SN1993J} \\
SN 1994I & 8.9 & \citet{SN1994I} & \citet{SN1994I}\\
SN 1998bw & 38 & \citet{SN1998bw} & \citet{SN1998bw}\\
SN 2003bg & 19.6 & \citet{SN2003bg} & \citet{SN2003bg} \\
SN 2003L & 92 & \citet{SN2003L} & \citet{SN2003L} \\
\hline
SN 2004cc & 18 & \\
SN 2004dk & 23 & \citet{SNe2004} & \citet{SNe2004}\\
SN 2004gq & 26 & \\
\hline
SN 2008ax & 9.6 & \citet{SN2008ax}  & \citet{SN2008axD} \\
SN 2008iz & 3.5 &  \citet{SN2008iz} & \citet{SN2008iz} \\
SN 2011dh & 8.4 & \citet{SN2011dh}  & \citet{SN2011dh} \\
\hline
\end{tabular}
\label{snsample}
\end{table*}

\begin{table*}
\centering
\caption{Gamma Ray Bursts used in the analysis.}
\begin{tabular}{cccc}
\hline
Source & Distance (Mpc) &  Reference & Distance reference \\
\hline
GRB 970508 & 5299 &  \citet{GRB970508} & \citet{GRB970508} \\
GRB 030329 & 802 & \citet{GRB030329} & \citet{GRB030329D} \\
GRB 060418 & 10917 & \citet{GRB060418} & \citet{GRB060418D} \\
GRB 110709B & 36523 & \citet{GRB110709B}  & \citet{GRB110709B}\\
\hline
\end{tabular}
\label{grbsample}
\end{table*}

\begin{table*}
\centering
\caption{Classical / Dwarf / Recurring Novae used in the analysis.}
\begin{tabular}{cccc}
\hline
Source & Distance (Mpc) &  Reference & Distance reference \\
\hline
RS Oph & 0.0016 & \citet{RSOph} & \citet{RSOph}\\
Sco 2012 & 0.0067 &  \citet{SCO2012}  & \citet{SCO2012} \\
SS Cyg & 0.000114 & \citet{SSCyg} & \citet{SSCygD} \\
T Pyx  & 0.0048 & \citet{TPyx} & \citet{TPyx}\\
V407 Cyg & 0.003 &  \citet{V407} & \citet{V407}\\
V1500 Cyg & 0.00135 & \citet{V1500Cyg} & \citet{V1500CygD}\\
V1723 Aql & 0.005 & \citet{V1723} & \citet{V1723}  \\
V1974 Cyg & 0.0018 & \citet{V1974Cyg} & \citet{V1974D} \\
\hline
\end{tabular}
\label{novasample}
\end{table*}

\begin{table*}
\centering
\caption{ A sample of X-ray binaries, including black hole X-ray binaries (BHXRB), neutron star X-ray binaries (NSXRB), binary pulsar and ultraluminous X-ray source (ULX) used in the analysis.}
\begin{tabular}{cccc}
\hline
Source & Distance (Mpc) &  Reference & Distance reference \\
\hline
Aql X-1 & 0.005 & GBI & \citet{aqld} \\
B1259-63& 0.0023 & \citet{B1259-63} & \citet{B1259-63D}\\
CI Cam & 0.01 & GBI & \citet{CICamD} \\
Cir X-1 & 0.0078 & \citet{CirX-1} & \citet{CirX1D} \\
Cyg X-1 & 0.00186 & GBI  & \citet{xrbd}\\
Cyg X-2 & 0.011 & GBI & \citet{aqld} \\
Cyg X-3 & 0.0072 & \citet{CygX-3}, GBI & \citet{CygX3d} \\ 
GRO J1655-40 & 0.0032 & \citet{J1655-40} & \citet{xrbd} \\
GRS 1915+105 & 0.011 & GBI & \citet{xrbd}\\
GX13+1 & 0.007 & \citet{GX13+1} & \citet{GX13+1}\\
GX17+2 & 0.0075 & GBI & \citet{GX17+2D} \\
GX 339-4 & 0.008 & \citet{GX339-4} & \citet{GX339-4D} \\
LS 5039& 0.0029 & \citet{LS5039} & \citet{LS5039d}\\
LSI+61$^\circ$303 & 0.002 & GBI & \citet{lsid}\\
MAXIJ1836-194 & 0.008 & \citet{MaxiJ1836-194}& \citet{MaxiJ1836-194D} \\
Sco X-1 & 0.0028 & GBI  & \citet{scodist1}\\
SGR 1806-20 & 0.009 & \citet{SGR1806-20} & \citet{MagnetarD} \\
SS 433 & 0.0055 & \citet{SS433}, GBI & \citet{SS433D} \\
XTE  J1550-564 & 0.00438 & \citet{XTEJ1550-564}   & \citet{XTEJ1550-564D} \\
XTE J1748−288 & 0.008 & \citet{XTE1748−288}  & \citet{XTEJ1748-288} \\
XTE J1859+226 & 0.0042 & \citet{XTE1748−288} & \citet{J1859+226D} \\
M31 ULX & 0.78 & \citet{M31ULX} & \citet{M31ULX} \\
\hline
\end{tabular}
\label{xrbsample}
\end{table*}

\begin{table*}
\centering
\caption{RSCVn used in the analysis.}
\begin{tabular}{cccc}
\hline
Source & Distance (Mpc) &  Reference & Distance reference \\
\hline
CF Oct & 0.000194 & \citet{CFOct} & \citet{CFOctD} \\
HR1099 & 0.00003068 & GBI & \citet{HR1099D}  \\
UX Ari & 0.00005 & GBI & \citet{UXAriD} \\
\hline
\end{tabular}
\label{rscvnsample}
\end{table*}

\begin{table*}
\centering
\caption{Algol systems used in the analysis.}
\begin{tabular}{cccc}
\hline
Source & Distance (Mpc) &  Reference & Distance reference \\
\hline
Algol & 0.000029 & GBI & \citet{UXAriD}\\
$\delta$ Lib & 0.000095 & GBI & \citet{deltalibD}\\
RZ Cas & 0.000063 & \citet{RZCas} & \citet{RZCas} \\
\hline
\end{tabular}
\label{algolsample}
\end{table*}

\begin{table*}
\centering
\caption{T Tauri included in the analysis.}
\begin{tabular}{cccc}
\hline
Source & Distance (Mpc) &  Reference & Distance reference \\
\hline
V773 Tau & 0.0001328 & \citet{TTauri} & \citet{TTauriD} \\
\hline
\end{tabular}
\label{ttaurisample}
\end{table*}

\begin{table*}
\centering
\caption{Magnetic CVs used in the analysis.}
\begin{tabular}{cccc}
\hline
Source & Distance (Mpc) &  Reference & Distance reference\\
\hline
AE Aqr  & 0.000102 & \citet{AEAqr}, \citet{AEAqr1} & \citet{AEAqrD} \\
V834 Cen & 0.000086 & \citet{V834Cen} & \citet{V834D} \\
\hline
\end{tabular}
\label{mcvsample}	
\end{table*}

\begin{table*}
\centering
\caption{Flare Stars used in the analysis.}
\begin{tabular}{cccc}
\hline
Source & Distance (Mpc) &  Reference & Distance reference\\
\hline
AD Leo & 0.0000047 & \citet{FS} &\citet{FS} \\
AU Mic & 0.00001 &\citet{FS} & \citet{FS} \\
UV Ceti & 0.0000026 & \citet{FS}&\citet{FS}\\
YZ CMi & 0.0000059  &\citet{FS} &\citet{FS}\\
\hline
\end{tabular}
\label{fssample}
\end{table*}


\begin{thebibliography}{}


\bibitem[\protect\citeauthoryear{Abada-Simon et 
al.}{1993}]{AEAqr1} Abada-Simon M., Lecacheux A., Bastian 
T.~S., Bookbinder J.~A., Dulk G.~A., 1993, ApJ, 406, 692 

\bibitem[\protect\citeauthoryear{Araujo-Betancor et 
al.}{2005}]{V834D} Araujo-Betancor S., G{\"a}nsicke B.~T., 
Long K.~S., Beuermann K., de Martino D., Sion E.~M., Szkody P., 2005, ApJ, 
622, 589 

\bibitem[\protect\citeauthoryear{Audard, Donisan, G{\"u}del}{2005}]{RZCas} Audard M., Donisan J.~R., G{\"u}del M., 2005, ESASP, 560, 407 

\bibitem[\protect\citeauthoryear{Bak{\i}{\c s} et 
al.}{2006}]{deltalibD} Bak{\i}{\c s} V., Budding E., Erdem A., 
Bak{\i}{\c s} H., Demircan O., Hadrava P., 2006, MNRAS, 370, 1935 

\bibitem[\protect\citeauthoryear{Bartlett et 
al.}{2013}]{CICamD} Bartlett E.~S., Clark J.~S., Coe M.~J., 
Garcia M.~R., Uttley P., 2013, MNRAS, 429, 1213 

\bibitem[\protect\citeauthoryear{Bastian, Dulk, 
\& Chanmugam}{1988}]{AEAqr} Bastian T.~S., Dulk G.~A., Chanmugam G., 1988, ApJ, 324, 431 

\bibitem[\protect\citeauthoryear{Bell et al.}{2011}]{NGC7213} 
Bell M.~E., et al., 2011, MNRAS, 411, 402 

\bibitem[\protect\citeauthoryear{Berger}{2003}]{2003AIPC..662..420B} Berger 
E., 2003, AIPC, 662, 420 

\bibitem[\protect\citeauthoryear{Blundell 
\& Bowler}{2004}]{SS433D} Blundell K.~M., Bowler M.~G., 2004, ApJ, 616, L159 

\bibitem[\protect\citeauthoryear{Borisova et 
al.}{2005}]{CFOctD} Borisova A.~P., Innis J.~L., Coates D.~W., 
Tsvetkov M.~K., Schr{\"o}der K.-P., Bues I., 2005, ESASP, 560, 453

\bibitem[\protect\citeauthoryear{Booth et al.}{2009}]{MeerKAT} 
Booth R.~S., de Blok W.~J.~G., Jonas J.~L., Fanaroff B., 2009, arXiv, 
arXiv:0910.2935 

\bibitem[\protect\citeauthoryear{Bradshaw, Fomalont, 
\& Geldzahler}{1999}]{scodist1} Bradshaw C.~F., Fomalont E.~B., Geldzahler B.~J., 1999, ApJ, 512, L121 

\bibitem[\protect\citeauthoryear{Britzen et 
al.}{2010}]{1803+784} Britzen S., et al., 2010, A\&A, 511, A57

\bibitem[\protect\citeauthoryear{Brocksopp et 
al.}{2007}]{XTEJ1748-288} Brocksopp C., Miller-Jones J.~C.~A., 
Fender R.~P., Stappers B.~W., 2007, MNRAS, 378, 1111 

\bibitem[\protect\citeauthoryear{Cao et al.}{2012}]{Cao2012} 
Cao Y., et al., 2012, ApJ, 752, 133 

\bibitem[\protect\citeauthoryear{Cenko et al.}{2010}]{GRB060418} 
Cenko S.~B., et al., 2010, ApJ, 711, 641 

\bibitem[\protect\citeauthoryear{Chen, Jiang, 
\& Zhang}{2001}]{NRAO150} Chen Y.-J., Jiang D.-R., Zhang F.-J., 2001, ChJAA, 1, 507 

\bibitem[\protect\citeauthoryear{Chevalier}{1998}]{1998ApJ...499..810C} 
Chevalier R.~A., 1998, ApJ, 499, 810 

\bibitem[\protect\citeauthoryear{Chevalier}{1982}]{SN2a} 
Chevalier R.~A., 1982a, ApJ, 259, 302 

\bibitem[\protect\citeauthoryear{Chevalier}{1982}]{SN2b} 
Chevalier R.~A., 1982b, ApJ, 259, L85 

\bibitem[\protect\citeauthoryear{Chevalier 
\& Li}{1999}]{GRB7} Chevalier R.~A., Li Z.-Y., 1999, ApJ, 520, L29 

\bibitem[\protect\citeauthoryear{Chomiuk et 
al.}{2012}]{V407} Chomiuk L., et al., 2012, ApJ, 761, 173 

\bibitem[\protect\citeauthoryear{Church et 
al.}{2012}]{GX17+2D} Church M.~J., Gibiec A., Ba{\l}uci{\'n}ska-Church M., Jackson N.~K., 2012, A\&A, 546, A35 

\bibitem[\protect\citeauthoryear{D'A{\`i} et 
al.}{2012}]{CirX1D} D'A{\`i} A., et al., 2012, A\&A, 543, A20 

\bibitem[\protect\citeauthoryear{Dhawan, Mioduszewski, 
\& Rupen}{2006}]{lsid} Dhawan V., Mioduszewski A., Rupen M., 2006, smqw.conf,  

\bibitem[\protect\citeauthoryear{Duerbeck}{1999}]{AEAqrD} 
Duerbeck H.~W., 1999, IBVS, 4731, 1 

\bibitem[\protect\citeauthoryear{Ellison et 
al.}{2006}]{GRB060418D} Ellison S.~L., et al., 2006, MNRAS, 372, 
L38 

\bibitem[\protect\citeauthoryear{Emmanoulopoulos et 
al.}{ 2013}]{NGC7213D} Emmanoulopoulos D., Papadakis I.~E., 
Nicastro F., McHardy I.~M., 2013, MNRAS, 429, 3439 

\bibitem[\protect\citeauthoryear{Farinelli et 
al.}{2013}]{J1859+226} Farinelli R., et al., 2013, MNRAS, 428, 
3295 

\bibitem[\protect\citeauthoryear{Fender}{2006}]{XRB2} Fender 
R., 2006, AIPC, 856, 23 

\bibitem[\protect\citeauthoryear{Fender 
\& Bell}{2011}]{2011BASI...39..315F} Fender R.~P., Bell M.~E., 2011, BASI, 39, 315 

\bibitem[\protect\citeauthoryear{Frail, Waxman, 
\& Kulkarni}{2000}]{GRB970508} Frail D.~A., Waxman E., Kulkarni S.~R., 2000, ApJ, 537, 191 

\bibitem[\protect\citeauthoryear{Gaensler et 
al.}{2005}]{SGR1806-20} Gaensler B.~M., et al., 2005, Natur, 434, 
1104 

\bibitem[\protect\citeauthoryear{Gallo et al.}{2006}]{2004-447} 
Gallo L.~C., et al., 2006, MNRAS, 370, 245 

\bibitem[\protect\citeauthoryear{Gallo et al.}{2004}]{GX339-4} 
Gallo E., Corbel S., Fender R.~P., Maccarone T.~J., Tzioumis A.~K., 2004, 
MNRAS, 347, L52 

\bibitem[\protect\citeauthoryear{Galloway et 
al.}{2008}]{aqld} Galloway D.~K., Muno M.~P., Hartman J.~M., 
Psaltis D., Chakrabarty D., 2008, ApJS, 179, 360 

\bibitem[\protect\citeauthoryear{Gandhi et al.}{2010}]{GX339-4D} 
Gandhi P., et al., 2010, MNRAS, 407, 2166 

\bibitem[\protect\citeauthoryear{Giannios 
\& Metzger}{2011}]{TDE} Giannios D., Metzger B.~D., 2011, MNRAS, 416, 2102 

\bibitem[\protect\citeauthoryear{Giroletti, Taylor, 
\& Giovannini}{2005}]{NGC4278} Giroletti M., Taylor G.~B., Giovannini G., 2005, ApJ, 622, 178 

\bibitem[\protect\citeauthoryear{Gupta et al.}{2012}]{2012MNRAS.425.1357G} 
Gupta A.~C., et al., 2012, MNRAS, 425, 1357 

\bibitem[\protect\citeauthoryear{Hannikainen et 
al.}{2001}]{XTEJ1550-564} Hannikainen D., et al., 2001, ESASP, 459, 
291 

\bibitem[\protect\citeauthoryear{Hannikainen et 
al.}{2000}]{J1655-40} Hannikainen D.~C., Hunstead R.~W., 
Campbell-Wilson D., Wu K., McKay D.~J., Smits D.~P., Sault R.~J., 2000, 
ApJ, 540, 521 

\bibitem[\protect\citeauthoryear{Hassall, Keane, 
\& Fender}{2013}]{2013MNRAS.436..371H} Hassall T.~E., Keane E.~F., Fender R.~P., 2013, MNRAS, 436, 371 

\bibitem[\protect\citeauthoryear{Hjellming}{1996}]{V1974Cyg} 
Hjellming R.~M., 1996, ASPC, 93, 174 

\bibitem[\protect\citeauthoryear{Hjellming 
\& Johnston}{1988}]{1988ApJ...328..600H} Hjellming R.~M., Johnston K.~J., 1988, ApJ, 328, 600 

\bibitem[\protect\citeauthoryear{Homan et al.}{2006}]{2006ApJ...642L.115H} 
Homan D.~C., et al., 2006, ApJ, 642, L115 

\bibitem[\protect\citeauthoryear{Homan et 
al.}{2004}]{GX13+1} Homan J., Wijnands R., Rupen M.~P., Fender R., Hjellming R.~M., di Salvo T., van der Klis M., 2004, A\&A, 418, 255 

\bibitem[\protect\citeauthoryear{Huenemoerder et 
al.}{2013}]{HR1099D} Huenemoerder D.~P., Phillips K.~J.~H., 
Sylwester J., Sylwester B., 2013, ApJ, 768, 135 

\bibitem[\protect\citeauthoryear{Jauncey et 
al.}{2003}]{2003ASPC..300..199J} Jauncey D.~L., Bignall H.~E., Lovell 
J.~E.~J., Kedziora-Chudczer L., Tzioumis A.~K., Macquart J.-P., Rickett 
B.~J., 2003, ASPC, 300, 199 

\bibitem[\protect\citeauthoryear{Johnston et 
al.}{1999}]{B1259-63} Johnston S., Manchester R.~N., McConnell 
D., Campbell-Wilson D., 1999, MNRAS, 302, 277 

\bibitem[\protect\citeauthoryear{Johnston et 
al.}{2007}]{ASKAPa} Johnston S., et al., 2007, PASA, 24, 174 

\bibitem[\protect\citeauthoryear{Johnston et 
al.}{2008}]{ASKAPb} Johnston S., et al., 2008, ExA, 22, 151 


\bibitem[\protect\citeauthoryear{Kato 
\& Hachisu}{2005}]{V1974D} Kato M., Hachisu I., 2005, ApJ, 633, L117 

\bibitem[\protect\citeauthoryear{Keane}{2013}]{2013IAUS..291..295K} Keane 
E.~F., 2013, IAUS, 291, 295 

\bibitem[\protect\citeauthoryear{Koay et 
al.}{2011}]{2011A&A...534L...1K} Koay J.~Y., Bignall H.~E., Macquart J.-P., Jauncey D.~L., Rickett B.~J., Lovell J.~E.~J., 2011, A\&A, 534, L1 

\bibitem[\protect\citeauthoryear{K{\"o}rding et 
al.}{2008}]{SSCyg} K{\"o}rding E., Rupen M., Knigge C., 
Fender R., Dhawan V., Templeton M., Muxlow T., 2008, Sci, 320, 1318 

\bibitem[\protect\citeauthoryear{Krauss et al.}{2011}]{V1723} 
Krauss M.~I., et al., 2011, ApJ, 739, L6 
 
\bibitem[\protect\citeauthoryear{Krauss et al.}{2012}]{SN2011dh} 
Krauss M.~I., et al., 2012, ApJ, 750, L40 

\bibitem[\protect\citeauthoryear{Kudryavtseva et 
al.}{2011}]{0605-085} Kudryavtseva N.~A., et al., 2011, A\&A, 526, A51 

\bibitem[\protect\citeauthoryear{Kudryavtseva 
\& Pyatunina}{2006}]{2223-052} Kudryavtseva N.~A., Pyatunina T.~B., 2006, ARep, 50, 1 

\bibitem[\protect\citeauthoryear{Kulkarni et 
al.}{1998}]{SN1998bw} Kulkarni S.~R., et al., 1998, Natur, 395, 
663 

\bibitem[\protect\citeauthoryear{Kumar 
\& Panaitescu}{2000}]{GRB6} Kumar P., Panaitescu A., 2000, ApJ, 541, L9 

\bibitem[\protect\citeauthoryear{Lazio et al.}{2001}]{2001ApJS..136..265L} 
Lazio T.~J.~W., Waltman E.~B., Ghigo F.~D., Fiedler R.~L., Foster R.~S., 
Johnston K.~J., 2001, ApJS, 136, 265 

\bibitem[\protect\citeauthoryear{Levan et al.}{2011}]{swiftD} 
Levan A.~J., et al., 2011, Sci, 333, 199 

\bibitem[\protect\citeauthoryear{Li 
\& Chevalier}{1999}]{GRB1} Li Z.-Y., Chevalier R.~A., 1999, ApJ, 526, 716 

\bibitem[\protect\citeauthoryear{Ling, Zhang, 
\& Tang}{2009}]{CygX3d} Ling Z., Zhang S.~N., Tang S., 2009, ApJ, 695, 1111 

\bibitem[\protect\citeauthoryear{Lister et al.}{2009}]{mojave} 
Lister M.~L., et al., 2009, AJ, 137, 3718 

\bibitem[\protect\citeauthoryear{Livio}{1999}]{jets1} Livio 
M., 1999, PhR, 311, 225 

\bibitem[\protect\citeauthoryear{Lorimer et 
al.}{2007}]{2007Sci...318..777L} Lorimer D.~R., Bailes M., McLaughlin 
M.~A., Narkevic D.~J., Crawford F., 2007, Sci, 318, 777 

\bibitem[\protect\citeauthoryear{Lu et al.}{2012}]{NRAO530} Lu 
J.-C., Wang J.-Y., An T., Lin J.-M., Qiu H.-B., 2012, RAA, 12, 643 

\bibitem[\protect\citeauthoryear{Lyne et al.}{2010}]{intpulsar} 
Lyne A., Hobbs G., Kramer M., Stairs I., Stappers B., 2010, Sci, 329, 408 

\bibitem[\protect\citeauthoryear{Marchili et 
al.}{2010}]{SN2008iz} Marchili N., et al., 2010, A\&A, 509, A47 

\bibitem[\protect\citeauthoryear{Marchili et 
al.}{2012}]{0954+658} Marchili N., Krichbaum T.~P., Liu X., Song H.-G., Gab{\'a}nyi K.~{\'E}., Fuhrmann L., Witzel A., Zensus J.~A., 2012, A\&A, 542, A121 

\bibitem[\protect\citeauthoryear{Massi, Menten, 
\& Neidh{\"o}fer}{2002}]{TTauri} Massi M., Menten K., Neidh{\"o}fer J., 2002, A\&A, 382, 152 

\bibitem[\protect\citeauthoryear{Mesler et al.}{2012}]{GRB030329} 
Mesler R.~A., Pihlstr{\"o}m Y.~M., Taylor G.~B., Granot J., 2012, ApJ, 759, 4 

\bibitem[\protect\citeauthoryear{Middleton et 
al.}{2013}]{M31ULX} Middleton M.~J., et al., 2013, Natur, 493, 
187 

\bibitem[\protect\citeauthoryear{Milisavljevic et 
al.}{2012}]{SN1980KD} Milisavljevic D., Fesen R.~A., Chevalier 
R.~A., Kirshner R.~P., Challis P., Turatto M., 2012, ApJ, 751, 25 

\bibitem[\protect\citeauthoryear{Miller-Jones}{2014}]{xrbd} 
Miller-Jones J.~C.~A., 2014, PASA, 31, 16 

\bibitem[\protect\citeauthoryear{Miller-Jones et 
al.}{2013}]{SSCygD} Miller-Jones J.~C.~A., Sivakoff G.~R., 
Knigge C., K{\"o}rding E.~G., Templeton M., Waagen E.~O., 2013, Sci, 340, 
950 

\bibitem[\protect\citeauthoryear{Mitra-Kraev et 
al.}{2005}]{FSAlfven} Mitra-Kraev U., Harra L.~K., Williams D.~R., Kraev E., 2005, A\&A, 436, 1041 


\bibitem[\protect\citeauthoryear{Mold{\'o}n et 
al.}{2012}]{LS5039d} Mold{\'o}n J., Rib{\'o} M., Paredes J.~M., Brisken W., Dhawan V., Kramer M., Lyne A.~G., Stappers B.~W., 2012, A\&A, 543, A26 

\bibitem[\protect\citeauthoryear{Molnar, Reid, 
\& Grindlay}{1988}]{CygX-3} Molnar L.~A., Reid M.~J., Grindlay J.~E., 1988, ApJ, 331, 494 

\bibitem[\protect\citeauthoryear{Nelson et al.}{2012}]{TPyx} 
Nelson T., et al., 2012, arXiv, arXiv:1211.3112 

\bibitem[\protect\citeauthoryear{O'Brien et 
al.}{2006}]{RSOph} O'Brien T., et al., 2006, evn..conf,  

\bibitem[\protect\citeauthoryear{Ofek et al.}{2014}]{Ofek2014} 
Ofek E.~O., et al., 2014, ApJ, 788, 154 

\bibitem[\protect\citeauthoryear{Pastorello et 
al.}{2008}]{SN2008axD} Pastorello A., et al., 2008, MNRAS, 389, 
955 

\bibitem[\protect\citeauthoryear{P{\'e}rez-Torres, Alberdi, 
\& Marcaide}{2001}]{SN1993J} P{\'e}rez-Torres M.~A., Alberdi A., Marcaide J.~M., 2001, A\&A, 374, 997 

\bibitem[\protect\citeauthoryear{Peterson et 
al.}{2011}]{UXAriD} Peterson W.~M., Mutel R.~L., Lestrade 
J.-F., G{\"u}del M., Goss W.~M., 2011, ApJ, 737, 104 

\bibitem[\protect\citeauthoryear{Pyatunina et 
al.}{2006}]{Pyatunina2006} Pyatunina T.~B., Kudryavtseva N.~A., 
Gabuzda D.~C., Jorstad S.~G., Aller M.~F., Aller H.~D., Ter{\"a}sranta H., 
2006, MNRAS, 373, 1470 

\bibitem[\protect\citeauthoryear{Rau et al.}{2009}]{Rau2009} 
Rau A., et al., 2009, PASP, 121, 1334 


\bibitem[\protect\citeauthoryear{Readhead}{1994}]{Tlim} 
Readhead A.~C.~S., 1994, ApJ, 426, 51 

\bibitem[\protect\citeauthoryear{Rib{\'o}, Paredes, 
\& Mart{\'{\i}}}{1999}]{LS5039} Rib{\'o} M., Paredes J.~M., Mart{\'{\i}} J., 1999, NewAR, 43, 545 

\bibitem[\protect\citeauthoryear{Rickett}{2007}]{2007RMxAC..27..129R} 
Rickett B.~J., 2007, RMxAC, 27, 129 

\bibitem[\protect\citeauthoryear{Rickett et 
al.}{1995}]{1995A&A...293..479R} Rickett B.~J., Quirrenbach A., Wegner R., Krichbaum T.~P., Witzel A., 1995, A\&A, 293, 479 

\bibitem[\protect\citeauthoryear{Rodriguez 
\& Prat}{2008}]{XTE1748−288} Rodriguez J., Prat L., 2008, int..work,  

\bibitem[\protect\citeauthoryear{Roming et al.}{2009}]{SN2008ax} 
Roming P.~W.~A., et al., 2009, ApJ, 704, L118 

\bibitem[\protect\citeauthoryear{Roy et al.}{2000}]{0235+164} 
Roy M., Papadakis I.~E., Ramos-Col{\'o}n E., Sambruna R., Tsinganos K., 
Papamastorakis J., Kafatos M., 2000, ApJ, 545, 758 

\bibitem[\protect\citeauthoryear{Roy et al.}{2012}]{V1500Cyg} 
Roy N., et al., 2012, BASI, 40, 293 

\bibitem[\protect\citeauthoryear{Russell et 
al.}{2013}]{MaxiJ1836-194D} Russell D.~M., et al., 2013, ApJ, 768, L35 

\bibitem[\protect\citeauthoryear{Ryder et al.}{2003}]{SN2001ig} 
Ryder S.~D., Sadler E., Subrahmanyan R., Weiler K.~W., Panagia N., 
Stockdale C., 2003, astro, arXiv:astro-ph/0309378 

\bibitem[\protect\citeauthoryear{Salvi et al.}{2002}]{IIIzw2} 
Salvi N.~J., et al., 2002, MNRAS, 335, 177 

\bibitem[\protect\citeauthoryear{Sari, Piran, 
\& Halpern}{1999}]{GRB2} Sari R., Piran T., Halpern J.~P., 1999, ApJ, 519, L17 

\bibitem[\protect\citeauthoryear{Schlegel 
\& Petre}{2006}]{SN1988ZD} Schlegel E.~M., Petre R., 2006, ApJ, 646, 378 

\bibitem[\protect\citeauthoryear{Sesar et al.}{2007}]{SDSS} 
Sesar B., et al., 2007, AJ, 134, 2236 

\bibitem[\protect\citeauthoryear{Shannon, Johnston, 
\& Manchester}{2014}]{B1259-63D} Shannon R.~M., Johnston S., Manchester R.~N., 2014, MNRAS, 437, 3255 

\bibitem[\protect\citeauthoryear{Shaposhnikov 
\& Titarchuk}{2009}]{J1859+226D} Shaposhnikov N., Titarchuk L., 2009, ApJ, 699, 453 

\bibitem[\protect\citeauthoryear{Slavin, O'Brien, 
\& Dunlop}{1995}]{V1500CygD} Slavin A.~J., O'Brien T.~J., Dunlop J.~S., 1995, MNRAS, 276, 353 

\bibitem[\protect\citeauthoryear{Slee et al.}{1987}]{CFOct} 
Slee O.~B., Nelson G.~J., Stewart R.~T., Wright A.~E., Innis J.~L., Ryan 
S.~G., Vaughan A.~E., 1987, MNRAS, 229, 659 

\bibitem[\protect\citeauthoryear{Smith, G{\"u}del, 
\& Audard}{2005}]{FS} Smith K., G{\"u}del M., Audard M., 2005, A\&A, 436, 241 

\bibitem[\protect\citeauthoryear{Soderberg et 
al.}{2006}]{SN2003bg} Soderberg A.~M., Chevalier R.~A., Kulkarni 
S.~R., Frail D.~A., 2006, ApJ, 651, 1005

\bibitem[\protect\citeauthoryear{Soderberg et 
al.}{2005}]{SN2003L} Soderberg A.~M., Kulkarni S.~R., Berger 
E., Chevalier R.~A., Frail D.~A., Fox D.~B., Walker R.~C., 2005, ApJ, 621, 
908 

\bibitem[\protect\citeauthoryear{Soleri et al.}{2009}]{CirX-1} 
Soleri P., Tudose V., Fender R., van der Klis M., Jonker P.~G., 2009, 
MNRAS, 399, 453 

\bibitem[\protect\citeauthoryear{Song 
\& Liu}{2007}]{2007ASPC..373..743S} Song H.-G., Liu X., 2007, ASPC, 373, 743 

\bibitem[\protect\citeauthoryear{Steiner et 
al.}{2011}]{XTEJ1550-564D} Steiner J.~F., et al., 2011, MNRAS, 416, 
941 

\bibitem[\protect\citeauthoryear{Tendulkar, Cameron, 
\& Kulkarni}{2012}]{MagnetarD} Tendulkar S.~P., Cameron P.~B., Kulkarni S.~R., 2012, ApJ, 761, 76 
\newpage
\bibitem[\protect\citeauthoryear{Thornton et 
al.}{2013}]{2013Sci...341...53T} Thornton D., et al., 2013, Sci, 341, 53 


\bibitem[\protect\citeauthoryear{Torres et al.}{2012}]{TTauriD} 
Torres R.~M., Loinard L., Mioduszewski A.~J., Boden A.~F., 
Franco-Hern{\'a}ndez R., Vlemmings W.~H.~T., Rodr{\'{\i}}guez L.~F., 2012, 
ApJ, 747, 18 

\bibitem[\protect\citeauthoryear{Trushkin, Nizhelskij, 
\& Bursov}{2008}]{SS433} Trushkin S.~A., Nizhelskij N.~A., Bursov N.~N., 2008, mqw..conf,  

\bibitem[\protect\citeauthoryear{Trushkin, Nizhelskij, 
\& Zhekanis}{2011}]{MaxiJ1836-194} Trushkin S.~A., Nizhelskij N.~A., Zhekanis G.~V., 2011, ATel, 3656, 1 

\bibitem[\protect\citeauthoryear{Usher et al.}{2013}]{NGC4278D} 
Usher C., Forbes D.~A., Spitler L.~R., Brodie J.~P., Romanowsky A.~J., 
Strader J., Woodley K.~A., 2013, MNRAS, 436, 1172 

\bibitem[\protect\citeauthoryear{van der Horst et 
al.}{2008}]{GRB030329D} van der Horst A.~J., et al., 2008, A\&A, 480, 35 

\bibitem[\protect\citeauthoryear{van der Laan}{1966}]{1966Natur.211.1131V} 
van der Laan H., 1966, Natur, 211, 1131 

\bibitem[\protect\citeauthoryear{van Dyk et 
al.}{1993}]{SN1988Z} van Dyk S.~D., Weiler K.~W., Sramek R.~A., 
Panagia N., 1993, ApJ, 419, L69 

\bibitem[\protect\citeauthoryear{van Haarlem et 
al.}{2013}]{LOFAR1} van Haarlem M.~P., et al., 2013, A\&A, 556, A2 

\bibitem[\protect\citeauthoryear{Vermeulen 
\& Cohen}{1994}]{1994ApJ...430..467V} Vermeulen R.~C., Cohen M.~H., 1994, ApJ, 430, 467 

\bibitem[\protect\citeauthoryear{Villata et 
al.}{2004}]{BLLac} Villata M., et al., 2004, A\&A, 424, 497 

\bibitem[\protect\citeauthoryear{Volvach et 
al.}{2010}]{3C} Volvach A.~E., Ryabov M.~I., Volvach 
L.~N., Suharev A.~I., Aller H.~D., Aller M.~F., 2010, AIPC, 1206, 360 

\bibitem[\protect\citeauthoryear{Weiler et al.}{2011}]{SN1994I} 
Weiler K.~W., Panagia N., Stockdale C., Rupen M., Sramek R.~A., Williams 
C.~L., 2011, ApJ, 740, 79 

\bibitem[\protect\citeauthoryear{Weiler et al.}{1992}]{SN1980K} 
Weiler K.~W., van Dyk S.~D., Panagia N., Sramek R.~A., 1992, ApJ, 398, 248 

\bibitem[\protect\citeauthoryear{Weiler et 
al.}{2002}]{SN1} Weiler K.~W., Panagia N., Montes M.~J., Sramek R.~A., 2002, ARA\&A, 40, 387 

\bibitem[\protect\citeauthoryear{Wellons, Soderberg, 
\& Chevalier}{2012}]{SNe2004} Wellons S., Soderberg A.~M., Chevalier R.~A., 2012, ApJ, 752, 17 

\bibitem[\protect\citeauthoryear{Weston et al.}{2013}]{SCO2012} 
Weston J.~H.~S., et al., 2013, arXiv, arXiv:1306.2265 

\bibitem[\protect\citeauthoryear{Wright et al.}{1988}]{V834Cen} 
Wright A.~E., Stewart R.~T., Nelson G.~J., Slee O.~B., Cropper M., 1988, 
MNRAS, 231, 319 

\bibitem[\protect\citeauthoryear{Zauderer et 
al.}{2011}]{J1644+57} Zauderer B.~A., et al., 2011, Natur, 476, 
425 

\bibitem[\protect\citeauthoryear{Zauderer et 
al.}{2013}]{GRB110709B} Zauderer B.~A., Berger E., Margutti R., 
Pooley G.~G., Sari R., Soderberg A.~M., Brunthaler A., Bietenholz M.~F., 
2013, ApJ, 767, 152 
\end{thebibliography}
\end{document}